\documentclass[twocolumn,showpacs,preprintnumbers,amsmath,amssymb]{revtex4}
\usepackage{graphicx}% Include figure files
\usepackage{dcolumn}% Align table columns on decimal point
\usepackage{bm}% bold math

\begin{document}

%\preprint{APS/123-QED}

\title{\boldmath
Difference in $B^+$ and $B^0$ Direct $CP$ Asymmetry as Effect of a
Fourth Generation }
%
%\vfill
\author{Wei-Shu Hou}
%Lines break automatically or can be forced with \\
\author{Makiko Nagashima}%
%\affiliation{%
%Authors' institution and/or address\\
%This line break forced with \textbackslash\textbackslash
%}%
%
\author{Andrea Soddu}
% \homepage{http://www.Second.institution.edu/~Charlie.Author}
%\affiliation{${}^a$Institute of Physics, Academia Sinica,
%                 Taipei, Taiwan 115, R.O.C.}
\affiliation{Department of Physics, National Taiwan
University, Taipei, Taiwan 106, R.O.C. }%
%

%\date{\today}
%
%\vskip -1cm
%
\vfill
\begin{abstract}
Direct CP violation in $B^0\to K^+\pi^-$ decay has emerged at
$-10\%$ level, but the asymmetry in $B^+\to K^+\pi^0$ mode is
consistent with zero. This difference points towards possible New
Physics in the electroweak penguin operator. We point out that a
sequential fourth generation, with sizable $V^*_{t^\prime
s}V_{t^\prime b}$ and near maximal phase, could be a natural
cause. We use the perturbative QCD factorization approach for
$B\to K\pi$ amplitudes. While the $B^0\to K^+\pi^-$ mode is
insensitive to $t^\prime$, we critically compare $t^\prime$
effects on direct CP violation in $B^+\to K^+\pi^0$ with $b\to
s\ell^+\ell^-$ and $B_s$ mixing. If the $K^+\pi^0$--$K^+\pi^-$
asymmetry difference persists, we predict $\sin2\Phi_{B_s}$ to be
negative.
\end{abstract}
%\keywords{Suggested keywords}%Use showkeys class option if keyword
                              %display desired
\maketitle

%\protect \textbackslash\textbackslash}

%Before August in 2004, the direct CP asymmetries of $B^0\to K^+\pi^-$
%(${\cal A}_{\rm CP}^{+-}$) and $B^+\to K^+\pi^0$ (${\cal A}_{\rm CP}^{+0}$)
%had been reported as $-0.09\pm 0.04$ and $-0.10\pm 0.08$, respectively \cite{PDG}.

%Just three years after $CP$ violation (CPV) is observed %~\cite{phi101}
%in $B_d$ mixing,
Direct $CP$ violation (DCPV) in $B^0\to K^+\pi^-$ decay has
recently been observed~\cite{AKpi1,AKpi2} at the B factories. The
combined asymmetry
%, defined with respect to quark flavor,
is ${\cal A}_{K\pi}=-0.114\pm0.020$.
%. The negative value
%
%
However, the asymmetry in $B^+\to K^+\pi^0$ decay is found to
be~\cite{AKpi2,AKpi0BaBar04}
%consistent with zero.
% Belle experiment found~\cite{AKpi2} the to be
%$+0.04\pm0.05\pm0.02$, hence consistent with zero, and is
%supported by the BaBar value~\cite{AKpi0BaBar04} of
%$+0.06\pm0.06\pm0.01$.
%The combined result of
${\cal A}_{K\pi^0}=+0.049\pm0.040$, which differs from ${\cal
A}_{K\pi}$ by
\begin{eqnarray}
{\cal A}_{K\pi^0}-{\cal A}_{K\pi}=+0.163\pm0.045,
 \label{diff}
\end{eqnarray}
with 3.6$\sigma$ significance. All existing models have predicted
${\cal A}_{K\pi^0} \sim {\cal A}_{K\pi}$, as this basically
follows from isospin symmetry. The large difference of
Eq.~(\ref{diff}), if it persists, could indicate isospin breaking
New Physics (NP), likely~\cite{Buras,Barger,Nandi,
Mishima,Wu,Baek} through the electroweak penguin (EWP) operator.

In this paper we point out a natural source for such EWP effects:
the existence of a 4th generation. The $t^\prime$ quark can modify
the EWP coefficients, but leave the strong and electromagnetic
penguin coefficients largely intact. Eq.~(\ref{diff}) can be
accounted for, provided that $m_{t^\prime} \sim 300$ GeV, and the
quark mixing elements $V^*_{t^\prime s}V_{t^\prime b}$ is not much
smaller than $V_{cb}$ and has near maximal $CP$ phase.
Independently, $b\to s\ell^+\ell^-$ and $B_s$ mixing constraints
can allow large $t^\prime$ effects only if~\cite{Arhrib} the
associated $CP$ phase is near maximal.

Precision electroweak data imply that $|m_{t^\prime} -
m_{b^\prime}|$ cannot be too large~\cite{PDG}. Unitarity of quark
mixing requires $|V_{ub^\prime}| < 0.08$~\cite{PDG}, while
constraining $V^*_{t^\prime s}V_{t^\prime b}$ is the subject of
this paper. Since $b\to d$ transitions (including $B_d$ mixing)
appear Standard Model (SM) like, we set $V_{t^\prime d} \sim 0$.
We thus decouple from $s\to d$ constraints such as $\epsilon_K$
and $K\to\pi\nu\nu$ as well~\cite{FHS}.

%\section{$B\to K\pi$ with 4th generation}

Adding a fourth generation modifies short distance coefficients.
%and enlarges the quark mixing matrix $V$.
Defining $\lambda_q=V_{qs}^\ast V_{qb}$, the effective Hamiltonian
relevant for $B\to K\pi$ can be written as
\begin{eqnarray}
H_{\rm eff} \propto %\frac{G_F}{\sqrt 2} \left[
 \lambda_u \left( C_1 O_1 + C_2 O_2 \right)
%\nonumber \\ & &
 + \sum_{i=3}^{10}(\lambda_c C^t_{i}
                  -\lambda_{t^\prime} \Delta\,C_i) O_{i},
%  \right],
 \label{Ham}
\end{eqnarray}
where $O_{1,2}$ are the tree operators, $\lambda_c C^t_{i}$ are
the usual SM penguin terms, and $-\lambda_{t^\prime} \Delta\,C_i$
with $\Delta\,C_i \equiv C^{t^\prime}_{i} - C^t_{i}$ is the 4th
generation effect.
We have used $\lambda_u+\lambda_c+\lambda_t+\lambda_{t'}=0$,
simplified by ignoring $|\lambda_u| \lesssim 10^{-3}$, such that
$\lambda_t \cong - \lambda_c -\lambda_{t'}$~\cite{HWS}. The
penguin coefficients $\lambda_t C^t_{i}+\lambda_{t^\prime}
C^{t^\prime}_{i}$ at scale $\mu$ are then put~\cite{Arhrib} in the
form of Eq.~(\ref{Ham}), which respect the SM limit for
$\lambda_{t^\prime} \to 0$ or $m_{t^\prime} \to m_t$.
Explicit forms for $C_i$ and $O_i$ can be found, for example, in
Ref.~\cite{BBL}.

%%%%
\begin{figure}[b!]
\smallskip  %\smallskip
\includegraphics[width=1.9in,height=1.25in,angle=0]{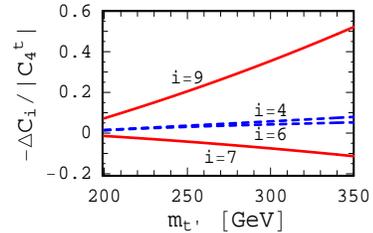}
 \vskip-0.2cm
%\smallskip\smallskip%\smallskip\smallskip\smallskip
 \caption{The $t^\prime$ correction $-\Delta\,C_i$ normalized to
 strength of strong penguin coefficient $|C_4^t|$ (both at $m_b$
 scale) vs. $m_{t^\prime}$.
 \label{fig:DeltaC}
}
\end{figure}
%%%%

The $K\pi$ amplitudes are dominated by $C_{4,6}^t$. To illustrate
${t^\prime}$ sensitivity, in Fig.~1 we plot $-\Delta\,C_i/|C_4^t|$
at $m_b$ scale vs $m_{t^\prime}$. The effect is clearly most
prominent for the EWP $C_9$ coefficient, with linear $x_{t^\prime}
\equiv m^2_{t^\prime}/M_W^2$ dependence arising from $Z$ and box
diagrams~\cite{HWS}. $\Delta C_7$ has similar dependence but has
weaker strength. For the strong penguin $\Delta C_{4,6}$, the
$t^\prime$ effect in the QCD penguin loop is weaker than
logarithmic~\cite{QCDP} and is very mild.
As we shall see, the $B^0 \to K^+\pi^-$ amplitude does not involve
the EWP. In contrast, the $B^+ \to K^+\pi^0$ amplitude is
sensitive to the EWP via $\Delta C_9 - \Delta C_7$ (virtual $Z$
materializing as $\pi^0$).

We see that it is natural for the 4th generation to show itself
through the EWP. The effect depends also on the quark mixing
matrix product, parameterized as~\cite{Arhrib}
\begin{eqnarray}
\lambda_{t^\prime} = V^*_{t^\prime s}V_{t^\prime b} = r_s\,
e^{i\phi_s}.
 \label{ltp}
\end{eqnarray}
%since it is in general complex. Its
The phase $\phi_s$ is needed to affect the CPV observables,
Eq.~(\ref{diff}). Most works on the 4th generation have ignored
the phase in $V^*_{t^\prime s}V_{t^\prime b}$, making the 4th
generation effect far less flexible nor interesting.

%% table for PQCD hard contributtions %%
\begin{table*}[t]
 \caption{\label{tab:PQCDkpiI}
  Factorizable contributions for $B^{0[+]}\to K^+\pi^{-[0]}$ in Standard
  Model, and for $m_{t'}=300$ GeV.
  The difference between the $t^\prime$ and $t$ penguin contributions
  gives $\Delta F_j^P$. ``N.A." stands for ``not applicable".
 }
\begin{ruledtabular}
\begin{tabular}{cccc}
& tree & $t$ penguin & $t'$ penguin\\ \hline $F^{(P)}_{e}$  &
$0.841$ [$0.843$]
               & $-0.074$ [$-0.075$]
               & $-0.076$ [$-0.078$] \\
$F^{(P)}_{a}$  & N.A. %\hskip2.15cm
 [$0.001+0.002\,i$]
               & $0.003+0.026\,i$ [$0.003+0.026\,i$]
               & $0.003+0.026\,i$ [$0.003+0.026\,i$] \\
$F^{(P)}_{ek}$ & N.A. %\hskip1.1cm
[$-0.105$]
               & N.A. %\hskip1.1cm
                [$-0.014$]
               & N.A. %\hskip1.1cm
                [$-0.029$] \\
\end{tabular}
\end{ruledtabular}
\end{table*}
%%

%\section{PQCD approach based on $k_T$ factorization}

Let us first see how ${\cal A}_{K\pi} < 0$ can be generated.
In the usual QCD factorization (QCDF) approach~\cite{BBNS}, strong
phases are power suppressed, while strong penguin $C_4$ and $C_6$
coefficients pick up perturbative absorptive parts. %~\cite{BSS}.
Thus, the predicted ${\cal A}_{K\pi}$ is small, and turns out to
be positive. For the perturbative QCD factorization
(PQCDF)~\cite{KLS} approach, one has an additional absorptive part
coming from the annihilation diagram, which arises from a cut on
the two quark lines in $B\to \bar sq \to K\pi$ decay. In this way,
the PQCDF approach predicted~\cite{KLS} the sign and order of
magnitude of ${\cal A}_{K\pi}$. By incorporating annihilation
contributions as in PQCDF, however, QCDF can also~\cite{BN} give
negative ${\cal A}_{K\pi}$.

We adopt PQCDF as a definite calculational framework. The
$\overline B^0\to K^-\pi^+$ amplitude for the 3 generation SM is
roughly given by
\begin{eqnarray}
{\cal M}_{K^-\pi^+}^{\rm SM} \ \propto \
 \lambda_u f_K F_e + \lambda_c ( f_K F_e^{P} + f_B F_a^{P}),
 \label{KpiSM}
\end{eqnarray}
where %(following PQCDF notation)
$F_e^{(P)}$ is the color-allowed tree (strong penguin)
contribution and is real, and $F_a^P$ is the strong penguin
annihilation term that has a large imaginary part. We have dropped
subdominant non-factorizable effects for sake of presentation.
Details cannot be given here, but these factorizable contributions
can be computed by following Ref.~\cite{KLS}, convoluting the hard
part (related to short distance coefficients $C_i$) and the soft,
nonperturbative meson wave functions. Basically, all the
%factorizable contributions
$F_j^{(P)}$s are integrals over Bessel functions, and in
particular, a Hankel function for $F_a^P$~\cite{KLS}. We give the
SM numbers for $F_e$, $F_e^P$ and $F_a^P$ in Table I, which leads
to ${\cal A}_{K\pi} =-0.16$ for $\phi_3 \equiv \arg \lambda_u^* =
60^\circ$ (value used throughout~\cite{phi3}), compared to the
experimental value of $-0.114\pm0.020$.

For $B^-\to K^-\pi^0$, the difference with $K^-\pi^+$ is
\begin{eqnarray}
 \sqrt2 {\cal M}_{K^-\pi^0}^{\rm SM} - {\cal M}_{K^-\pi^+}^{\rm SM}
  \ \propto \ \lambda_u f_{\pi}F_{ek} +\lambda_c f_{\pi} F_{ek}^{P},
 \label{Kpi0SM}
\end{eqnarray}
where $F_{ek}$ is the color suppressed tree term, while
$F_{ek}^{P}$ is the color allowed EWP, and both are real. A
negligible tree annihilation term $\lambda_u f_B F_a$ has been
dropped. Since both the $F_{ek}$ and $F_{ek}^{P}$ terms are
subdominant compared to $F_{e}^{P}$ in the 3 generation SM, ${\cal
A}_{K\pi^0}$ and ${\cal A}_{K\pi}$ cannot be far apart. From the
values of $F_{ek}$ and $F_{ek}^P$ given in Table~I, we get ${\cal
A}_{K\pi^0} = -0.10$, which is less negative than ${\cal
A}_{K\pi}$, but at some variance with Eq.~(\ref{diff}).

Adding the $t^\prime$ quark, one finds ${\cal M}_{K^-\pi^+} \cong
{\cal M}_{K^-\pi^+}^{\rm SM}$. The difference is proportional to
$\lambda_{t'}(f_K \Delta F_e^{P} +f_B \Delta F_a^{P})$, which is
small unless $\lambda_{t'}$ is very large. This is because
$F_{e,a}^{P}$ are strong penguins, hence $\Delta F_{e,a}^{P}$
depends very weakly on $m_{t^\prime}$, as can be seen from Table I
(for $m_{t^\prime} = 300$ GeV) and Fig.~1. Thus, ${\cal A}_{K\pi}$
is insensitive to the 4th generation.
For $K^-\pi^0$, one finds
\begin{eqnarray}
 \sqrt 2 {\cal M}_{K^-\pi^0} - \sqrt 2 {\cal M}_{K^-\pi^0}^{\rm SM}
 \ \propto %
  -\lambda_{t'} %(f_K \Delta F_e^{P} +f_B \Delta F_a^{P}
% \nonumber \\
%  && \hspace{3mm}+
f_{\pi}\Delta F_{ek}^{P},
 \label{Kpi0tp}
\end{eqnarray}
where again $\Delta F_{e,a}^{P}$ terms have been dropped, and
$\Delta F_{ek}^{P}$ is the $t^\prime$ correction to the EWP, which
is generated by $\Delta C_9 - \Delta C_7$ at short distance.
%From the NP phase in Eq.~(\ref{ltp}) and the expected strength of
%$\Delta F_{ek}^{P}$ for large $m_{t^\prime}$ (see Table I for
%$m_{t^\prime} = 300$ GeV case), we see that the 4th generation can
%indeed impact on ${\cal A}_{K\pi^0}$.

Let us put the $K^-\pi^+$ and $K^-\pi^0$ amplitudes in more
heuristic form. Eq.~(\ref{KpiSM}) can be put in the form
\begin{eqnarray}
{\cal M}_{K^-\pi^+} \approx {\cal M}_{K^-\pi^+}^{\rm SM}
 \ \propto \ r e^{-i\phi_3} + e^{i\delta},
 \label{Kpiphases}
\end{eqnarray}
%where $\phi_3 = \arg \lambda_u^*$,
and the 4th generation effect is minor. The ratio $r =
|\lambda_u|f_K F_e/\lambda_c|f_K F_e^P+f_B F_a^P|$ parameterizes
the relative strength of tree ($T$)  vs. strong penguins ($P$),
and $\delta$ is the strong phase of $f_K F_e^P+f_B F_a^P$ arising
from $F_a^{P} \equiv |F_a^P|e^{i\delta_a}$.
Analogously, for $K^-\pi^0$ one roughly has
\begin{eqnarray}
{\cal M}_{K^-\pi^0} &\propto&
   r\left(1+\frac{f_{\pi}F_{ek}}{f_K F_e}\right)e^{-i\phi_3}
 +\frac{f_{\pi}F_{ek}^P}{|f_K F_e^P+f_B F_a^P|}
 \nonumber\\
&& \hspace{-1.5mm}+ e^{i\delta} -\frac{f_{\pi}\Delta
F_{ek}^P}{|f_K F_e^P+f_B F_a^P|}
   \left\vert\frac{V^*_{t^\prime s}V_{t^\prime b}}
                  {V^*_{cs}V_{cb}}\right\vert e^{i\phi_s},
 \label{Kpi0phases}
\end{eqnarray}
where $F_{ek}$ and $F_{ek}^P$ terms come from SM (see
Eq.~(\ref{Kpi0SM})), and the $\Delta F_{ek}^P$ term comes from the
$t^\prime$ effect of Eq.~(\ref{Kpi0tp}). Since $r \sim 1/5$, we
see from Table I that, for $m_{t^\prime} \sim 300$ GeV and
$|V_{t^\prime s}V_{t^\prime b}| \equiv r_s$ not much smaller than
$|V_{cb}| \sim 0.04$, the impact of $t^\prime$ on ${\cal
A}_{K\pi^0}$ could be significant.

\begin{figure}[b!]
%\smallskip  %\smallskip
\includegraphics[width=1.6in,height=1.15in,angle=0]{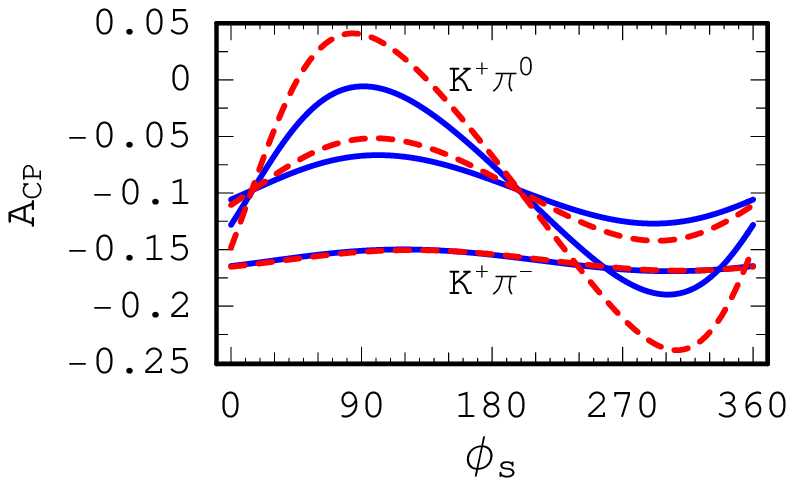}
\includegraphics[width=1.6in,height=1.15in,angle=0]{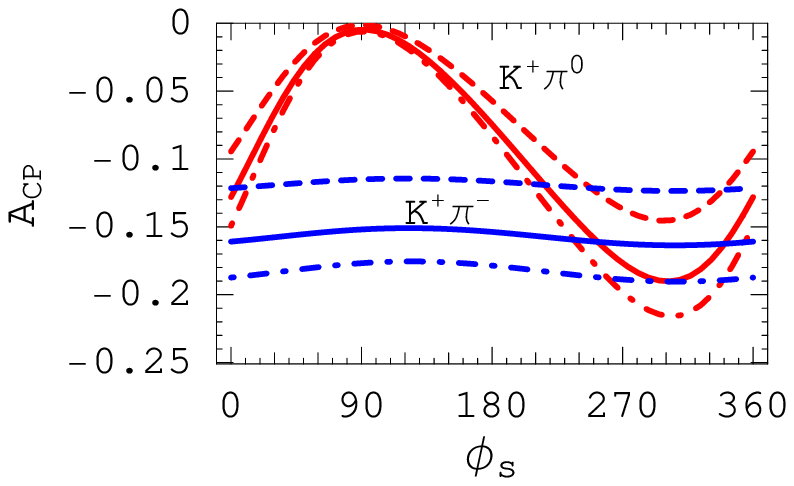}
%\smallskip\smallskip%\smallskip\smallskip\smallskip
 \caption{Direct CPV asymmetries
  ${\cal A}_{K\pi}$ and ${\cal A}_{K\pi^0}$
   vs. $\phi_s \equiv \arg V^*_{t^\prime s}V_{t^\prime b}$.
  In (a), the solid and dashed curves are for $m_{t^\prime} = 300$
  and 350 GeV, respectively, and for
  $r_s \equiv \vert V^*_{t^\prime s}V_{t^\prime b}\vert = 0.01$ and $0.03$.
  All curves for ${\cal A}_{K\pi}$ coalesce, but for ${\cal A}_{K\pi^0}$,
  the $r_s=0,03$ curves are steeper.
  For (b), the strong penguin absorptive phase $\delta$ is varied from
   $155^\circ$ (dotdash), $156^\circ$ (solid) to $160^\circ$ (dash)
   for $m_{t^\prime} = 300$ GeV and $r_s = 0.03$.
 }
 \label{fig:Fig2}
\end{figure}

We have presented in the above the major contributions in PQCDF
framework. Performing a detailed calculation following
Ref.~\cite{KLS}, we plot ${\cal A}_{K\pi}$ and ${\cal A}_{K\pi^0}$
in Fig.~2(a) for $m_{t^\prime} = 300$, 350 GeV and $r_s = 0.01$
and 0.03. We see that, indeed, ${\cal A}_{K\pi}$ is almost
independent of $t^\prime$, while it is clear that the largest
impact on ${\cal A}_{K\pi^0}$ is for $\phi_s \sim \pm\pi/2$ and
large $m_{t^\prime}$ and $r_s$. To maximize ${\cal A}_{K\pi^0} -
{\cal A}_{K\pi} > 0$, $\phi_s \sim +\pi/2$ is selected, and
Eq.~(\ref{diff}) can in principle be accounted for.

The ${\cal A}_{K\pi} \sim -0.16$ value is at some variance with
the experimental value of $-0.114\pm0.020$. This number depends
crucially on the strong penguin phase. Rather than varying
detailed model parameters, we vary $\delta \equiv \arg\, (f_K
F_e^P+f_B F_a^P)$.
% of Eqs.~(\ref{Kpiphases}) and (\ref{Kpi0phases}) around
%the nominal value of $156^\circ$.
The sign difference between tree and strong penguin constitutes a
phase of $\pi$, and $\pi - \delta \sim 24^\circ$ is perturbative.
We plot ${\cal A}_{K\pi}$ and ${\cal A}_{K\pi^0}$ vs. $\phi_s$ in
Fig.~2(b) for $m_{t^\prime} = 300$ GeV and $r_s = 0.03$, for
$\delta = 155^\circ$, $156^\circ$ (nominal) and $160^\circ$. We
see that a slightly smaller $\pi - \delta$ lowers
$|{\cal A}_{K\pi}|$ %(vanishes with $\pi - \delta \rightarrow 0$)
and is preferred. Note that ${\cal A}_{K\pi^0} \sim 0$ around
$\phi_s \sim 90^\circ$ is due to a near cancellation between the
$\phi_3$ (tree) and $\phi_s$ (EWP) contributions.
Thus, we think PQCDF can account for ${\cal A}_{K\pi} =
-0.114\pm0.020$ without affecting ${\cal A}_{K\pi^0}$, but the NP
phase $\phi_s$ should be rather close to $90^\circ$.

%\section{constraint from $B\to X_s \ell\ell$ and $B_s$ mixing}

To entertain a large EWP effect in CPV in $b\to s$ decay, one
needs to be mindful of the closely related $b\to s\ell^+\ell^-$
and $B_s$ mixing constraints, as well as the usually stringent
$b\to s\gamma$ constraint.
%It was pointed out in Ref.~\cite{Arhrib} that these bounds are
%most tolerant when $V^*_{t^\prime s}V_{t^\prime b}$ is
%close to imaginary. But now that $B\to X_s\ell^+\ell^-$ rate
%is measured, we perform a detailed analysis.
%
We have checked that the $b\to s\gamma$ rate constraint is well
satisfied for the range of parameters under discussion. This is
because on-shell photon radiation is generated by the $b\to s$
transition operator $O_{7\gamma}$, and the associated coefficient
$\Delta C_{7\gamma}$ has weaker $m_{t^\prime}$ dependence than
$\Delta C_7$ shown in Fig.~1.
However, $b\to s\ell^+\ell^-$ is generated by EWP~\cite{HWS}
operators very similar to $O_{7-10}$ in Eq.~(\ref{Ham}) for $b\to
s\bar qq$. The difference is basically just in the $Z$ charge of
$q$ vs. $\ell$, hence with same $m_{t^\prime}$ dependence. The box
diagram for ${B_s}$ mixing also has similar $m_{t^\prime}$
dependence. Taking the formulas from Ref.~\cite{Arhrib}, we plot
$b\to s\ell^+\ell^-$ rate ($m_{\ell\ell} > 0.2$ GeV) and $\Delta
m_{B_s}$ vs. $\phi_s$ in Figs. 3(a) and (b), for $m_{t^\prime} =
300$, 350 GeV and $r_s = 0.01$ %, 0.02
 and 0.03.
%The CPV phase in $B_s$ mixing, $\sin2\Phi_{B_s}$,
%is plotted in Fig.~2(d). Its value can be predicted for
%given $m_{t^\prime}$, $r_s$ and $\phi_s$.

%
\begin{figure}[b!]
\smallskip  %\smallskip
\hspace{3mm}\vspace{-2.mm}
\includegraphics[width=1.59in,height=1.0in,angle=0]{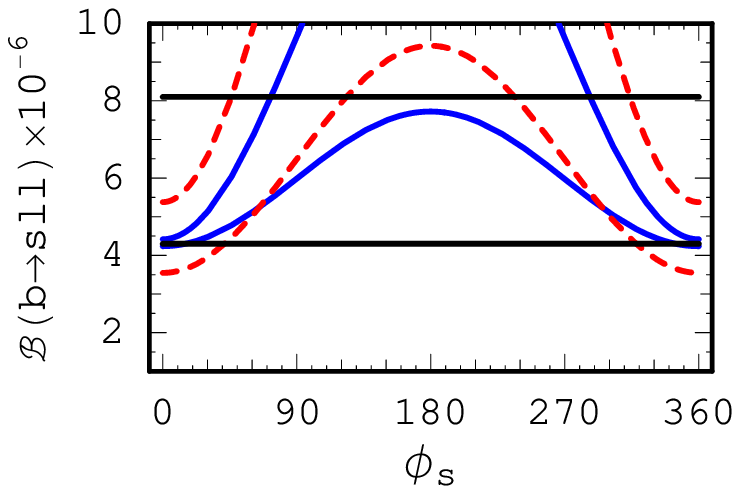}
\hspace{-1.5mm}\vspace{-2.mm}
\includegraphics[width=1.6in,height=1.0in,angle=0]{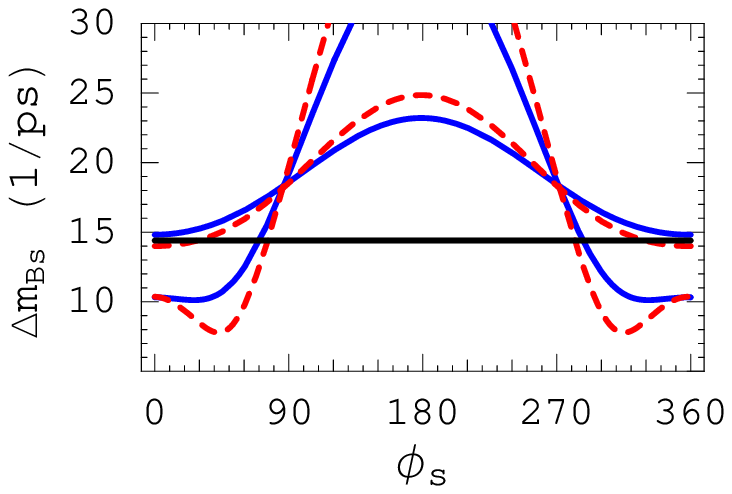}
\vspace{3mm} \hspace{2.5mm}
\includegraphics[width=1.62in,height=1.0in,angle=0]{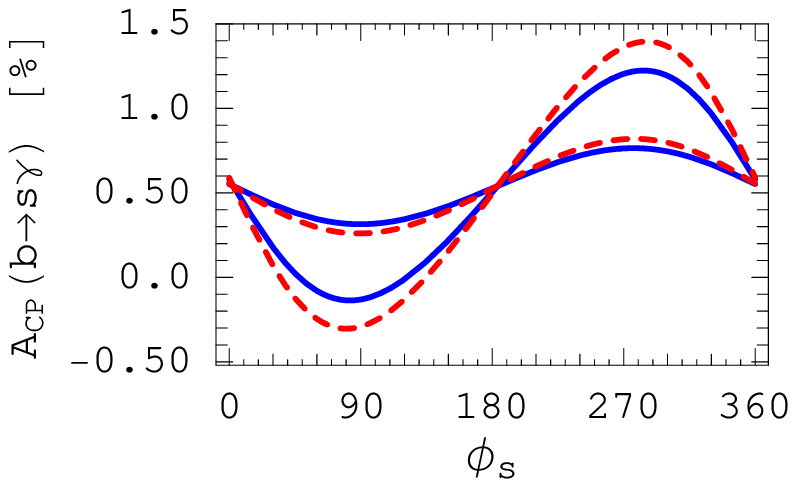}
\vspace{3mm}\hspace{-1mm}
\includegraphics[width=1.61in,height=1.0in,angle=0]{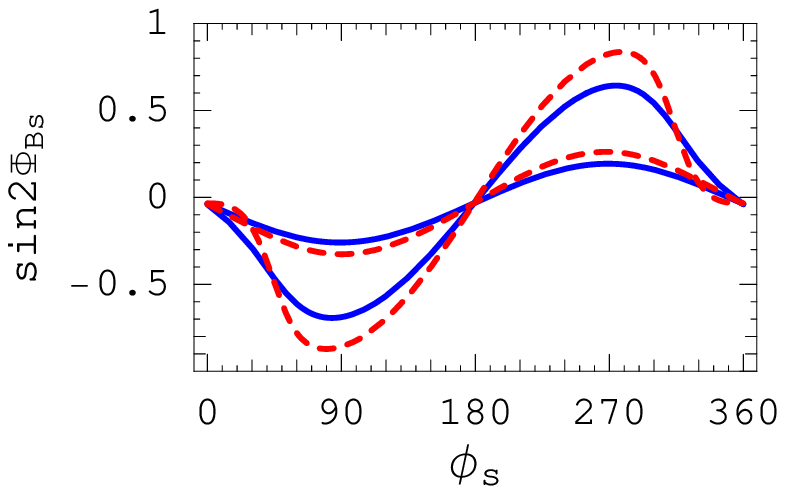}
\vspace{-1mm}
 %\smallskip\smallskip%\smallskip\smallskip\smallskip
 \vskip-0.1cm
\caption{
  (a) ${\cal B}(b\to s\ell^+\ell^-)$,
  (b) $\Delta m_{B_s}$,
  (c) $A_{\rm CP}(b\to s\gamma)$ and
  (d) $\sin 2\Phi_{B_s}$
   vs. $\phi_s = \arg V^*_{t^\prime s}V_{t^\prime b}$.
  Notation is same as Fig.~2(a), with effect strongest for larger
  $r_s$ and $m_{t^\prime}$.
  Horizontal solid band in (a) corresponds to $1\sigma$ experimental
  range, and solid line in (b) is the lower limit,
  both from Ref.~\cite{PDG}. The experimental range for (c) is
  outside the plot.
  }
 \label{fig:Fig3}
%\vspace{-2mm}
\end{figure}

We can understand the finding of Ref.~\cite{Arhrib} that
$\phi_s\sim 90^\circ$ is best tolerated by the $b\to
s\ell^+\ell^-$ and $\Delta m_{B_s}$ constraints.
%The $t^\prime$ effect is clearly very strong in both observables.
%
For $\cos\phi_s < 0$, the $b\to s\ell^+\ell^-$ rate gets greatly
enhanced~\cite{HWS}, and would run against recent measurements.
One is therefore forced to the $\cos\phi_s > 0$ region, where
$t^\prime$ effect is destructive against SM $t$ effect.
For $\Delta m_{B_s}$, the effect gets destructive for $\cos\phi_s
> 0$ when $r_s$ is sizable. Since one just has a lower
bound~\cite{PDG} of 14.4 ps$^{-1}$, $\Delta m_{B_s}$ tends to push
one away from the $\cos\phi_s > 0$ region. The combined effect is
to settle around $\phi_s \sim \pm\pi/2$, i.e.
imaginary~\cite{Arhrib}. This result is independent of the
discrepancy of Eq.~(\ref{diff}).

For sake of discussion we have plotted, as horizontal solid
straight lines in Fig.~3(a), the 1$\sigma$ range of ${\cal B}(B\to
X_s\ell^+\ell^-) = (6.1^{+2.0}_{-1.8})\times 10^{-6}$~\cite{PDG}
for $m_{\ell\ell} > 0.2$ GeV. This is the Particle Data Group
(PDG) 2004 average over Belle and BaBar
results~\cite{BelleXsll03,BaBarXsll04}, with a combined total of
154M $B\overline B$ pairs.
Belle has recently measured~\cite{BelleXsll} with 152M $B\overline
B$ pairs the value ${\cal B}(B\to X_s\ell^+\ell^-) =
(4.11\pm0.83^{+0.74}_{-0.70})\times 10^{-6}$ for $m_{\ell\ell} >
0.2$ GeV, which would be more stringent. However, this lower
result should be confirmed by BaBar, hence we use the more
conservative~\cite{Xsll} PDG 2004 range.
%
%, feeding the combined HFAG~\cite{HFAG} summer 2004 result
%of $(4.46^{+0.98}_{-0.96})\times 10^{-6}$. We plot (dash lines)
%the $2\sigma$ range in Fig.~2(a).
%
For $\Delta m_{B_s}$, we plot the PDG bound of 14.4
ps$^{-1}$~\cite{PDG} as horizontal solid straight line in
Fig.~3(b).

Comparing Figs.~2(a) and 3(a), 3(b), we set ${\cal A}_{K\pi^0} >
-0.05$ as a requirement for a solution, for otherwise it is hard
to satisfy Eq.~(\ref{diff}), and in any case the 4th generation
would seem no longer needed. This requirement demands $r_s >
0.01$. For $m_{t^\prime} = 350$ GeV and $r_s = 0.03$, which can
best bring ${\cal A}_{K\pi^0} \gtrsim 0$, Figs.~3(a) and 3(b)
mutually exclude each other.
For $m_{t^\prime} = 300$ GeV and $r_s = 0.03$ (the case for
$m_{t^\prime} = 350$ GeV and $r_s = 0.02$ is very similar), one
finds $\phi_s \simeq 75^\circ$ gives ${\cal A}_{K\pi^0} \sim 0$.
However, ${\cal B}(b\to s\ell\ell)$ must be close to the maximal
value of $\sim 8\times 10^{-6}$, and $\Delta m_{B_s}$ would be
just above the bound.
%This case may seem somewhat unlikely in view of the new
%Belle result~\cite{BelleXsll} of
%$(4.11\pm0.83^{+0.74}_{-0.70})\times 10^{-6}$ for $B\to
%X_s\ell\ell$.
%
For lower $r_s$ values, the solution space is broader. For
example, for $m_{t^\prime} = 300$ GeV and $r_s = 0.02$, one has
${\cal A}_{K\pi^0} \gtrsim -0.05$ for $\phi_s \sim
63^\circ$--$100^\circ$. ${\cal B}(b\to s\ell\ell)$ can reach below
$6\times 10^{-6}$, but then $\Delta m_{B_s}$ would again approach
the current bound.

We see that for a range of parameter space roughly around
$m_{t^\prime} \sim 300$ GeV and $0.01 < r_s \lesssim 0.03$,
solutions to Eq.~(\ref{diff}) can be found that do not upset $b\to
s\ell\ell$ and $\Delta m_{B_s}$. Both large $t^\prime$ mass and
sizable $V_{t^\prime s}$ mixing are needed; no solutions are found
for $m_{t^\prime}=250$ GeV.

As the CPV effect through the EWP is large, one may worry if
similar effects may show up already in $b\to s\gamma$. We follow
Ref.~\cite{KN}, extend to 4 generations, and plot $A_{\rm CP}(b\to
s\gamma)$ vs $\phi_s$ in Fig.~3(c). Like the $A_{K\pi^0}$ case,
the $t^\prime$ effect cancels against the SM phase. $\vert A_{\rm
CP}(b\to s\gamma)\vert$ is in general {\it smaller} than the SM
value of $\sim 0.5\%$, and consistent with the current measurement
of $0.004\pm0.036$~\cite{HFAG}. In fact, it is below the
sensitivity for the proposed high luminosity ``Super B factory".

As prediction, we find $\sin2\Phi_{B_s} < 0$ for CPV in $B_s$
mixing, which is plotted vs $\phi_s$ in Fig.~3(d). We find
$\sin2\Phi_{B_s}$ in the range of $-0.2$ to $-0.7$ and correlating
with ${\cal A}_{K\pi^0}-{\cal A}_{K\pi}$. Three generation SM
predicts zero. Note that refined measurements of ${\cal B}(b\to
s\ell\ell)$ and future measurements of $\Delta m_{B_s}$ and
$\sin2\Phi_{B_s}$, together with theory improvements, can pinpoint
$m_{t^\prime}$, $r_s$ and $\phi_s$.
We note further that~\cite{PDG} 14.4 ps$^{-1} < \Delta m_{B_s} <$
21.8 ps$^{-1}$ cannot yet be excluded because data is compatible
with a signal in this region. We eagerly await $B_s$ mixing and
associated CPV measurement in the near future.

It is of interest to predict the asymmetries for the other two
$B\to K\pi$ modes. $K^0\pi^-$ is analogous to ${\cal
M}_{K^-\pi^+}$ except tree contribution is absent. We find ${\cal
M}_{\overline K^0\pi^-} \cong {\cal M}^{\rm SM}_{\overline
K^0\pi^-} \propto \lambda_c(f_K F_e^P + f_B F_a^P)$, so ${\cal
A}_{K^0\pi} \simeq 0$ and insensitive to $t^\prime$.
For $\overline B^0\to \overline K^0\pi^0$, we have ${\cal
M}_{\overline K^0\pi^0} \propto \lambda_u f_\pi F_{ek} + \lambda_c
( - f_K F_e^P - f_B F_a^P + f_\pi F_{ek}^P) - \lambda_{t^\prime}
f_\pi \Delta F_{ek}^P$.
% is similar to Eq.~(\ref{Kpi0phases}), with the first $e^{-i\phi_3}$
%term $\to 0$, $F_e^P \to - F_e^P$, and $r$, $\delta \to r^\prime$,
%$\delta^\prime$ from normalization change, but
Numerics can still be obtained from Table~I, giving ${\cal
A}_{K^0\pi} - {\cal A}_{K^0\pi^0} \sim 0.1$ if ${\cal A}_{K\pi^0}
- {\cal A}_{K\pi}$ is of order suggested by Eq.~(\ref{diff}).
The impact on mixing-dependent CPV in $\phi K_S$ and
$\eta^\prime K_S$ modes are %at the few \% level and
insignificant~\cite{Arhrib}.
%, but may lower $S_{\overline K^0\pi^0}$ to $\sim$ 0.5--0.6.
%within experimental errors, and in any case,
%Belle and BaBar do not yet agree on these modes.

The measurement of ${\cal A}_{K\pi^0}$ itself should not yet be
viewed as settled, since the recent BaBar value of
$+0.06\pm0.06\pm0.01$ changed sign from the
previous~\cite{AKpi0BaBar03} value of $-0.09\pm0.09\pm0.01$. But
if ${\cal A}_{K\pi^0}\sim 0$ hence Eq.~(\ref{diff}) stays, we
would need a large effect in the EWP with a new CPV phase. Note
that, unlike most treatments of the EWP, our strong phase is not a
fitted parameter, but calculated from PQCDF~\cite{Fek}.

%Earlier discussions~\cite{Buras} of possible NP in EWP focused on
%the ratios
%
%$R = \overline\Gamma(K\pi)/\overline\Gamma(K^0\pi)$,
%
%$R_c = 2\overline\Gamma(K\pi^0)/\overline\Gamma(K^0\pi)$,
%
%and $R_n = \overline\Gamma(K\pi)/2\overline\Gamma(K^0\pi^0)$. Here
%$K^{(0)}$, $\pi^{(0)}$ stand for charged (neutral) kaon and pion,
%and overline means averaging over $B$ and $\overline B$. The
%current experimental values~\cite{HFAG} are $R = 0.82\pm0.06$,
%$R_c = 1.00\pm0.08$ and $R_n = 0.79\pm0.08$. Both $R$ and $R_c$
%have moved down by more than $1\sigma$ compared to previous
%result. Calculating ${\cal M}_{\overline K^0\pi^-}$ and ${\cal
%M}_{\overline K^0\pi^0}$ in PQCDF, the three generation SM result
%is $R\cong 0.84$, $R_c\cong 1.11$ and $R_n\cong 0.67$. $R$ is
%insensitive to $t^\prime$, and $R \simeq 0.84$ is in good
%agreement with experiment. We plot $R_c$ and $R_n$ in Fig.~3(a)
%vs. $\phi_s$ for $m_{t^\prime}$ and $r_s$ values as before. We see
%that the $t^\prime$ effect improves agreement with experiment for
%$\phi_s < 90^\circ$, and data seems to favor a stronger effect
%such as $m_{t^\prime} = 300$ (350) GeV and $r_s = 0.03$ (0.02).
%
%We note, however, that experiment has yet to stabilize.

We have also studied separately the final state rescattering (FSI)
model~\cite{CHY} as a different proposed source of strong phase.
In this model, one allows $K^+\pi^{-,0} \leftrightarrow
K^0\pi^{0,+} \leftrightarrow K^{0,+}\eta$
%and $K^+\pi^0 \leftrightarrow K^0\pi^+ \leftrightarrow K^+\eta$
rescattering in the final state (power suppressed in QCDF and
PQCDF), and, to avoid double counting, one uses naive
factorization amplitudes as source before rescattering. In this
way, one can account~\cite{CHY} for ${\cal A}_{K\pi} < 0$, and
also generate a sizable $\pi^0\pi^0$ via rescattering from
$\pi^+\pi^-$. Neither QCDF nor PQCDF can account for ${\cal B}(B^0
\to \pi^0\pi^0) > 10^{-6}$. However, in contrast to
Eq.~(\ref{diff}), ${\cal A}_{K\pi^0}$ is found~\cite{CHY} to be
more negative than ${\cal A}_{K\pi}$ for ${\cal A}_{K\pi} < 0$. We
find no solution to Eq.~(\ref{diff}), even when $t^\prime$ is
considered. Besides the problem that already exists in 3
generation SM, rescattering brings the electroweak penguin into
the $K^-\pi^+$ amplitude from the $\overline K^0\pi^0$ mode, so
adding the $t^\prime$ does not help.

%The authors  of Ref.~\cite{Buras} emphasize $R_c$ and $R_n$, and
%make use of SU(3) symmetry to relate to $\pi\pi$ modes. We have
%not focused on $R_c$ and $R_n$, in part because the experiments
%have yet to settle, and also because our $t^\prime$ effect helps
%reduce the problem. We also refrain from making connection with
%$B\to \pi\pi$ decay. When considering NP, clearly one should
%not~\cite{Baek} assume SU(3), as for example we have taken
%$V_{t^\prime d} \sim 0$, which decouples us from $b\to d$ (and
%$s\to d$) transitions. The quark mixing pattern manifestly
%violates SU(3).

We have shown that a fourth generation $t^\prime$ quark can
account for ${\cal A}_{K\pi^0}\sim 0$. Using PQCD factorization
calculations, one can account for ${\cal A}_{K\pi} < 0$ (untouched
by $t^\prime$) and generate the needed ${\cal A}_{K\pi^0}-{\cal
A}_{K\pi}$ splitting, which repeats in ${\cal A}_{K^0\pi}-{\cal
A}_{K^0\pi^0}$. The closely related $b\to
s\ell^+\ell^-$ mode %, generated by the electroweak penguin,
should have rate not less than $6\times 10^{-6}$, and $B_s$ mixing
should not be far above the current bound of 14.4 ps$^{-1}$.
In fact, between the $b\to s \ell^+ \ell^-$ rate and the bound on
${B_s}$ mixing, $V^*_{t^\prime s}V_{t^\prime b}$ should be near
imaginary if one wants a large $t^\prime$ effect.
We predict a quite measurable $CP$ violating phase
$\sin2\Phi_{B_s}$ in the $-0.2$ to $-0.7$ range. Refined
measurements of the last three measurables can determine
$m_{t^\prime}$ and the strength and phase of $V^*_{t^\prime
s}V_{t^\prime b}$.

\vskip 0.3cm
\noindent{\bf Acknowledgement}.\ This work is
supported in part by NSC-93-2112-M-002-020, NSC93-2811-M-002-053
and NSC93-2811-M-002-047. We thank H.n.~Li for providing the
program for computing amplitudes in PQCDF.


\begin{thebibliography}{99}
\bibitem{phi3}
A $\pm 10^\circ$ change in $\phi_3$ hardly affects our results.
\end{thebibliography}
\end{document}